# Iridium 5$d$-electron driven superconductivity in ThIr$_3$

Karolina Górnicka[1], Debarchan Das[2], Sylwia Gutowska[3], Bartłomiej Wiendlocha[3], Michał J. Winiarski[1], Tomasz Klimczuk[1], Dariusz Kaczorowski[2,*]

[1]*Faculty of Applied Physics and Mathematics, Gdansk University of Technology, ul. Narutowicza 11/12, 80-233 Gdańsk, Poland*

[2]*Institute of Low Temperature and Structure Research, Polish Academy of Sciences, P.O.Box 1410, 50-590 Wrocław 2, Poland*

[3]*Faculty of Physics and Applied Computer Science, AGH University of Science and Technology, Aleja Mickiewicza 30, 30-059 Kraków, Poland*

**Abstract**

Polycrystalline sample of superconducting ThIr$_3$ was obtained by arc-melting Th and Ir metals. Powder x-ray diffraction revealed that the compound crystalizes in a rhombohedral crystal structure (*R-3m*, s.g. #166) with the lattice parameters: $a$ = 5.3414(2) Å and $c$ = 26.432(1) Å. Normal and superconducting states were studied by magnetic susceptibility, electrical resistivity and heat capacity measurements. The results showed that ThIr$_3$ is a type II superconductor (Ginzburg-Landau parameter $\kappa$ = 33) with the critical temperature $T_c$ = 4.41 K. The heat capacity data yielded the Sommerfeld coefficient $\gamma$ = 17.6 mJ mol$^{-1}$ K$^{-2}$ and the Debye temperature $\Theta_D$ = 169 K. The ratio $\Delta C/(\gamma T_c)$ = 1.6, where $\Delta C$ stands for the specific heat jump at $T_c$, and the electron–phonon coupling constant $\lambda_{e-p}$ = 0.74 suggest that ThIr$_3$ is a moderate-strength superconductor. The experimental studies were supplemented by band structure calculations, which indicated that the superconductivity in ThIr$_3$ is governed mainly by 5$d$ states of iridium. The significantly smaller band-structure value of Sommerfeld coefficient as well as the experimentally observed quadratic temperature dependence of resistivity and enhanced magnetic susceptibility suggest presence of electronic interactions in the system, which compete with superconductivity.

* Corresponding author: d.kaczorowski@intibs.pl

# 1. Introduction

Since the discovery of superconductivity over a century ago, the investigation of superconductors is still one of the most attractive topics in the condensed matter physics. Despite the fact that several distinct superconducting families have been found and studied thoroughly in hope of understanding various Cooper pairing mechanisms, new findings still emerge even from simple, binary intermetallic systems. In this respect, materials bearing $d$- and $f$- electrons are particularly intriguing, since their electronic properties are often dominated by strong spin-orbit coupling [1]. Recently, a homologous series $RE_{2m+n}T_{4m+5n}$, where RE is a light lanthanide metal, and T is a transition metal Ir or Rh has attracted a lot of attention [2–8]. The latter compounds constitute an important class of materials exhibiting a variety of physical phenomena, like superconductivity and different types of magnetic ordering [5–12]. $RE_{2m+n}T_{4m+5n}$ compounds crystallize with a centrosymmetric rhombohedral structure (space group $R\text{-}3m$) of the $PuNi_3$-type, build of $m$ $MgCu_2$-type blocks and $n$ $CaCu_5$-type blocks. Amidst $m = n = 1$ representatives, superconductivity was found in $LaIr_3$ and $CeIr_3$ below $T_c$ of 3.5 K and 2.5 K, respectively [5–7]. In the latter compound, Ce ions exhibit a strongly intermediate valence character. In turn, in $NdIr_3$ a ferromagnetic ground state was observed with the Curie temperature $T_C = 10.6$ K [8].

Recently, we succeeded in synthesizing the actinoid-bearing representative of the $RE_{2m+n}T_{4m+5n}$ family, viz. $ThIr_3$, which was briefly communicated by Geballe *et al.* [9] to be a superconductor with $T_c = 4.71$ K. Though the latter report was published over 50 years ago, to the best of our knowledge, no detailed studies have been performed as yet, aimed at determining the superconducting parameters of that compound.

In this paper, we report on the preparation of polycrystalline sample of $ThIr_3$ and our exhaustive investigation of its electronic properties by means of dc magnetization, electrical resistivity and heat capacity measurements. The experimental data are supplemented by the results of electronic band structure calculations.

## 2. Experimental and calculation details

Polycrystalline sample of $ThIr_3$ was prepared by arc-melting stoichiometric mixture of pure elements (purity: Th- 99.7 wt.%, Ir - 99.9 wt.%) under high-purity argon atmosphere using an arc furnace installed inside a Braun glove-box with controlled oxygen and moisture contents. To promote homogeneity, the button was turned over and remelted several times. Final mass loss was less than 1%. The product was stable against air and moisture, hard, and grey in color.

The crystal structure and phase purity of the sample was checked at room temperature by powder X-ray diffraction (pXRD) using a PANalytical diffractometer equipped with Cu K$\alpha$ radiation. The data were analyzed by the Rietveld method implemented in the FullProf package [13]. The crystal structure was drawn using the VESTA software [14].

A Quantum Design Dynacool Physical Property Measurement System (PPMS) with a vibrating sample magnetometer (VSM) and a Quantum Design Magnetic Property Measurement System (MPMS-XL) were used to measure the magnetic properties of the sample. The temperature dependencies of the zero-field-cooled (ZFC) and field-cooled (FC) magnetic susceptibility (defined as $dM/dH$, where $M$ is the magnetization and $H$ is the applied field strength) were measured in an external magnetic field of 20 Oe. Furthermore, the magnetic field variations of the ZFC magnetization were measured at various temperatures in the superconducting state. In addition, the temperature dependence of $dM/dH$ was determined in the normal state at temperatures up to 400 K in a magnetic field of 5 kOe. Electrical resistivity and heat capacity measurements were carried out in the temperature interval 0.4 – 300 K using a Quantum Design PPMS platform equipped with a Helium-3 refrigerator.

Temperature- and magnetic field-dependent electrical transport measurements were made using a standard four-probe technique, in which Pt wires were attached to the polished sample using conductive silver epoxy Epotek H20E. For the heat capacity measurements, a standard relaxation technique was used. The data were collected in zero magnetic field and magnetic fields up to 60 kOe.

Electronic structure calculations were undertaken to study the origin of electronic states building the Fermi surface of ThIr$_3$, to investigate the behavior of 5f states of Th and to calculate the band-structure values of the Sommerfeld coefficient and the Pauli paramagnetic susceptibility. We used the Quantum Espresso package [12, 13], based on pseudopotential and plane-waves methods, with Rappe-Rabe-Kaxiras-Joannopoulos (RRKJ) ultrasoft pseudopotentials and the Perdew-Burke-Ernzerhof generalized gradient approximation (PBE-GGA) for the exchange-correlation potential [17]. Wave function and charge density cutoff energies were set to 60 and 600 Ry, respectively. The lattice parameters and the atomic positions were optimized using the Broyden–Fletcher–Goldfarb–Shanno (BFGS) algorithm implemented in the Quantum Espresso package. The theoretical values of the rhombohedral lattice parameters *a* and *c* were found slightly smaller than the experimental ones, namely by -0.28 % and -0.75 %, respectively. To investigate the relativistic effects, two types of calculations were performed: scalar-relativistic and fully-relativistic with spin orbit coupling (SOC) effect considered for both Th and Ir atoms.

## 3. Results and discussion

### 3.1. Crystal structure

The pXRD pattern of ThIr$_3$ together with its Rietveld fitting profile and the Bragg positions are shown in **Figure 1**. The data pointed out rather good quality of the sample examined, although a small amount of ThO$_2$ (less than 1% wt.) was detected as an impurity.

An anisotropic broadening of reflections was observed, with the (00$l$) reflections being noticeably broader than expected. An anisotropic strain model was introduced in the refinement, which led to a significant increase of the fit quality ($\chi^2$ reduced from 11 to 6.5).

The crystallographic and refinement parameters are listed in **Table 1**. The calculations indicated that the compound crystallizes with a rhombohedral crystal structure (space group *R-3m*, No.166) of the PuNi$_3$–type with the lattice parameters: $a$ = 5.3394(1) Å and $c$ = 26.4228(8) Å. The refined values are larger than those reported for RE-based REIr$_3$ compounds, i.e. $a$ = 5.32 Å and $c$ = 26.34 Å for LaIr$_3$ [5], and $a$ = 5.2945(1) Å and $c$ = 26.219(1) Å for CeIr$_3$ [7].

The crystallographic unit cell of ThIr$_3$ is shown in the inset to **Figure 1**. It contains three nonequivalent positions of Ir atoms and two of Th atoms. Atomic environment of each of the sites is a Frank-Kasper polyhedron [18]. In one of its positions, Th is coordinated by sixteen Ir atoms, and in the other one by twelve Ir atoms forming truncated tetrahedrons.

One can distinguish three types of layers in the crystal structure of ThIr$_3$, namely the plane consisting of Th(3$a$) and Ir(6$c$) atoms [with distances Th(3$a$)-Ir(6$c$) = Ir(6$c$)-Ir(6$c$) = 3.08 Å], the plane of Ir(18$h$) atoms [with Ir(18$h$)-Ir(18$h$) distance of 2.67 Å) and a sandwich made of three planes of Th(6$c$), Ir(3$b$) and Th(3$a$) atoms [with Th(3$a$)-Ir(3$b$) distance equal to 3.17 Å]. The observed anisotropic PXRD peak broadening can likely be attributed to the presence of some disorder in the layer stacking sequence.

The nearest neighbors in the ThIr$_3$ unit cell are Ir(18$h$) atoms with Ir(18$h$) atoms and Ir(18$h$) atoms with Ir(6$c$) atoms located at the same distance of 2.67 Å. Slightly larger distance (2.71 Å) is between Ir(3$b$) and Ir(18$h$) atoms. As can be inferred from **Figure 1**, two types of polyhedra with Ir(18$h$) atoms in the corners are formed, one is centered at Ir(6$c$) atom and the other one is centered at Ir(3$b$) atom. In turn, Th(3$a$) atom has six Ir(6$c$) atoms as its nearest neighbors located at a distance of 3.08 Å. Similarly, Th(6$c$) atom has six Ir(18$h$) atoms as the nearest neighbors with Th(6$c$)-Ir(18$h$) distance equal to 3.06 Å. In both cases Ir atoms are organized into hexagons, in-plane and out-of-plane, for Th(3$a$) and Th(6$c$) environment, respectively (see **Figure 1**).

*3.2. Superconducting properties*

The main panel of **Figure 2** shows the low-temperature magnetic susceptibility ($\chi$) of ThIr$_3$ measured in a small applied field of 20 Oe in the ZFC and FC regimes. Strong diamagnetic signal observed in the ZFC data corroborates the superconducting ground state, first reported in Ref. 9. The superconducting critical temperature, estimated as an intersection between the line set by the steepest slope of the superconducting signal and the line obtained by extrapolation of the normal state magnetic susceptibility to lower temperatures [19], equals $T_c$ = 4.41 K. In turn, defining $T_c$ as an onset of diamagnetic ZFC susceptibility yields a value of 4.5 K, which is closer to the literature data ($T_c$ = 4.71 K [9,20]). At 2 K, the diamagnetic susceptibility corrected for the demagnetization factor $N$ = 0.55 (obtained from the $M(H)$ fit discussed next) amounts to about -0.7 x (1/4$\pi$). In contrast to the ZFC data, the FC diamagnetic susceptibility measured in the superconducting state is very small. This finding implies substantial pinning effect, likely at grain boundaries in the polycrystalline sample studied. The inset to **Figure 2** presents the magnetic field dependence of the magnetization measured in the superconducting state of ThIr$_3$ with increasing and decreasing the magnetic field strength. It is apparent from the figure that the compound exhibits a conventional type-II superconductivity. Above $T_c$, ThIr$_3$ is a Pauli-type paramagnet with the total magnetic susceptibility of about 40 × 10$^{-5}$ emu/mol (not shown).

In order to determine the lower critical field of ThIr$_3$, the magnetization was measured as a function of magnetic field at several temperatures below the superconducting transition temperature $T_c$ (see **Figure 3(a)**). For each temperature, the experimental data obtained in small magnetic fields were fitted using the proportionality $M_{fit} = -pH$, appropriate for a full shielding effect. Comparing the value of prefactor $p$ derived from the isotherm taken at $T = 1.7$ K with the ideal diamagnetism quantified as $\frac{-1}{4\pi}$, the demagnetization factor $N = 0.55$ was found, fairly consistent with shape of the sample used in the magnetic measurements. In the next step, from the plot of the difference $M-M_{fit}$ versus $H$ (see **Figure 3(b)**), the values of the lower critical field were extracted for each temperature (note the black dashed line), as displayed in **Figure 3(c)**. The so-obtained data points were analyzed with the expression:

$$H_{c1}(T) = H_{c1}(0)\left[1 - \left(\frac{T}{T_c}\right)^2\right] \qquad (1)$$

that yielded the parameters and $T_c = 4.65(3)$ K and $H_{c1}(0) = 27(1)$ Oe. Correcting the latter value for the demagnetization effect, the lower critical field $H_{c1}(0) = 60$ Oe was determined. Remarkably, the critical temperature obtained from this analysis is very close to $T_c$ derived from the magnetic susceptibility measurements, as well as the values derived from the electrical resistivity and heat capacity data (see below), thus proving the correctness of the approach applied.

The superconducting transition for $ThIr_3$ was further characterized through temperature and magnetic field dependent measurements of the electrical resistivity. As can be inferred from the main panel of **Figure 4(a)**, the normal-state resistivity reveals metallic behavior ($d\rho/dT > 0$), although the residual resistivity ratio RRR= $\rho(300K)/\rho(5K) = 1.6$ is rather small. The latter feature can be attributed to polycrystalline nature of the sample investigated that probably contained many macroscopic defects. The observed strongly curvilinear character of $\rho(T)$ may originate from sizable contribution of Mott-type interband scattering processes to the electrical conduction in $ThIr_3$ [21–23]. The onset of the superconducting state manifests itself as a sharp drop in the resistivity down to zero. The critical temperature defined as a 90% decrease of the resistivity with respect to its normal state value amounts to $T_c = 4.4$ K, in good agreement with the magnetic susceptibility data. In applied magnetic fields, the width of the superconducting transition slightly increases and $T_c$ systematically shifts to lower temperature with increasing the field strength (see the inset to **Figure 4(a)**). Applying the same criterion as for the zero-field $\rho(T)$ data, one can derive the temperature variation of the upper critical field ($\mu_0 H_{c2}$), shown in **Figure 4(b)**.

According to the Werthamer-Helfand-Hohenberg (WHH) approach [24,25], the orbital upper critical field at 0 K in a single-band type-II BCS superconductor can be estimated from the formula

$$H_{c2}(0) = -AT_c \frac{dH_{c2}}{dT}\bigg|_{T=T_c} \tag{2}$$

where A is the purity factor given by 0.693 for the dirty and 0.73 for the clean limit. For ThIr$_3$ one finds $dH_{c2}/dT$ = -15.4(1) kOe/K (note the red straight line in **Figure 4(b)**), which implies $H_{c2}(0)$ = 47 and 49 kOe in the dirty and clean limit scenario, respectively. The obtained values are distinctly smaller than the Pauli limiting field $H_{c2}^p(0) = 1.85T_c = 82$ kOe, calculated for ThIr$_3$ assuming weak electron-phonon coupling.

Then, if one assumes the upper critical field to be purely orbital, the coherence length can be derived from the Ginzburg-Landau formula $H_{c2} = \Phi_0/2\pi\xi_{GL}^2$, where $\Phi_0 = hc/2e$ is the quantum flux. This way, we found for ThIr$_3$ the value $\xi_{GL}$= 83 Å within the dirty limit scenario. In the next step, from the relation

$$H_{c1} = \frac{\Phi_0}{4\pi\lambda_{GL}^2} \ln\frac{\lambda_{GL}}{\xi_{GL}}, \tag{3}$$

we estimated the superconducting penetration depth $\lambda_{GL}$ = 3150 Å. The so-obtained parameters yielded the Ginzburg-Landau parameter $\kappa_{GL}= \lambda_{GL}/\xi_{GL}$ = 38 corroborating that ThIr$_3$ is a type-II superconductor. Finally, from the relationship

$$H_{c1}H_{c2} = H_c^2 \ln\kappa_{GL} \tag{4}$$

one can determine the thermodynamic critical field in the studied compound to be $H_c$ = 884 Oe.

**Figure 4(c)** displays $\rho(T)$ of ThIr$_3$ measured in applied magnetic field of 40 kOe. These data, representing the electrical resistivity in the normal state, can be approximated by the function

$$\rho(T) = \rho_0 + AT^2 , \qquad (5)$$

where the first term is the residual resistivity due to crystal defects and the second one accounts for electron-electron scattering processes [26,27]. The least-squares fitting of Eq. 5 to the experimental data in the range 3-7 K yielded the parameters: $\rho_0 = 164.86(1)$ µΩcm and $A = 0.01854(3)$ µΩcm/K$^2$.

The results of low-temperature heat capacity measurements are summarized in **Figure 5**. The zero-field raw data displayed in the panels (a) and (c) were corrected for the impurity contribution due to 1% wt. of ThO$_2$, based on the specific heat data of the latter material, reported by Magnani et al. [28]. The pronounced sharp anomaly in $C/T(T)$ confirms bulk nature of the superconductivity in ThIr$_3$. From the equal entropy construction shown in **Figure 5(a)**, which reflects the expected entropy balance between the normal state and the superconducting state at the superconducting phase transition, one finds the critical temperature $T_c = 4.41$ K, which is almost identical to those determined from the magnetic and resistivity data. The so-determined specific heat jump at $T_c$ is about $\Delta C = 124$ mJ/(molK).

In an external magnetic field of 60 kOe, the superconductivity in ThIr$_3$ is entirely suppressed (see **Figure 3(c)**), and thus the experimental data shown in **Figure 5(b)** represents the normal state of the compound. From the standard Debye formula $C/T = \gamma_n + \beta T^2$, applied in the temperature range up to 4 K, one obtains the Sommerfeld coefficient $\gamma_n = 19.4(4)$ mJ/(molK$^2$) and the parameter $\beta = 1.59(5)$ mJ/(molK$^4$), which gives the Debye temperature $\Theta_D = 169(2)$ K through the relationship

$$\Theta_D = \left(\frac{12\pi^4}{5\beta}nR\right)^{1/3} \qquad (6)$$

where R = 8.314 J/(molK) is the gas constant and $n$ = 4 is the number of atoms per formula unit.

The dimensionless electron-phonon coupling constant, which is a measure of the strength of attractive interaction between electrons and phonons, can be estimated from the inverted McMillan equation [29]

$$\lambda_{e-p} = \frac{1.04 + \mu^* \ln(\Theta_D/1.45 T_c)}{(1 - 0.62\mu^*)\ln(\Theta_D/1.45 T_c) - 1.04} \qquad (7)$$

where $\mu^*$ is the repulsive screened Coulomb part, usually set to $\mu^*$ = 0.13 for intermetallic superconductors [30,31]. In the case of ThIr$_3$, one obtains $\lambda_{e-p}$ = 0.74 indicating a moderately coupled superconductor.

**Figure 5(c)** shows the electronic specific heat in the superconducting state up to 2 K, derived by subtracting the lattice $\beta T^3$ contribution from the $C(T)$ data. The solid red line in that figure represent a least-squares fitting to the experimental results of the function $C_{el}(T)$ = $\gamma_{imp}T + B \cdot \exp(-\Delta/k_B T)$, which considers some contribution from a fraction of non-superconducting impurity phase (the first term) in addition to the standard fully gapped superconductivity (the exponential term, where k$_B$ stands for the Boltzmann constant). The analysis yielded $\gamma_{imp}$ = 1.8 mJ/(mol K$^2$) and $\Delta$ = 0.50(5) meV. It is worthwhile noting that the so-obtained energy gap is somewhat smaller than the estimate $\Delta$ = 1.76k$_B T_c$ = 0.67 meV provided by the BCS theory.

By subtracting $\gamma_{imp}$ from the Sommerfeld coefficient $\gamma$ obtained in the normal state (see above) one can determine the intrinsic electronic specific heat coefficient in ThIr$_3$ to be $\gamma$ = 17.6 mJ/(molK$^2$). Using this value and the afore-derived specific heat jump at $T_c$, another important superconducting parameter can be calculated namely the ratio $\Delta C/\gamma T_c$ = 1.6. The obtained value is larger than the expected value of 1.43 for a weakly coupled BCS superconductor, thus supporting the finding from the McMillan approach (see above). All the superconducting and normal state parameters of ThIr$_3$ are gathered in **Table 2**.

*3.3. Electronic band structure*

The density of states (DOS) in ThIr$_3$ calculated without and with inclusion of the spin-orbit coupling is presented in **Figure 6**. In each case, the main contribution to DOS comes from Ir atoms, although Th atoms contribution is also significant, especially above the Fermi level. The overall shape of DOS is not strongly affected by SOC, however some important differences are seen near the Fermi level. As shown in details in **Figure 7**, $N(E_F)$ computed with SOC is considerably larger than that without SOC (3.46 eV$^{-1}$/f.u. and 2.45 eV$^{-1}$/f.u., respectively). The small peak, which appears in DOS in the full-relativistic calculations, originates from 5$f$ states of the Th(3$a$) atoms, as can be inferred from **Figure 8**, where partial DOS, decomposed over atomic orbitals, is plotted for each Th and Ir state (the partial densities at the Fermi level are listed in **Table 3**). As expected, the main contributions to the overall atomic densities come from Ir-5$d$ states and Th-6$d$ states, while Th- 5$f$ states contribute to the large DOS peak above $E_F$. The largest contribution to $N(E_F)$ comes from the Ir(6$c$) and Th(3$a$) atoms, which form together a layer in the crystal structure of ThIr$_3$ (see the inset to **Figure 1** and the discussion in Chapter 3.1).This effect is related to the spin-orbit interaction, as SOC significantly increases DOS values for both types of atoms.

In **Table 3,** one can find the electric charges derived by integrating partial DOS up to $E_F$. Iridium, with valence electron configuration 5$d^7$6$s^2$, and thorium, with valence configuration 6$d^2$7$s^2$, are expected to contribute nine and four electrons to the main valence band, respectively. For both elements, chemical bonding and hybridization effect may cause charge transfer from $s$ to $d$ states, and charge transfer from Ir atoms to Th atoms. Some electric charge is also transferred to Th-5$f$ orbitals, which are basically empty in elemental Th, yet are well known to exhibit much aptitude to hybridize with *s* and *d* states [32]. Depending on the experimental or computational technique, filling of the 5*f* orbital in crystalline Th atom was reported in the literature to span between 0.5 and 1.3 [29]. In the case of ThIr$_3$, the 5*f* orbital filling was estimated to be about 0.8 per atom. However, these states are of itinerant nature, with equal spin-up and spin-down occupation, hence no magnetic moment, associated with 5*f* electrons, is observed, in agreement with the experimental findings. **Figure 8** shows valence charge density plots, where charge transfers are clearly visible.

**Figure 9** displays the Fermi surface (FS) and the electronic dispersion relations in ThIr$_3$ calculated without SOC and with SOC (FS was visualized using the program XCrySDen [33]). Due to large number of bands (there are 93 valence electrons per unit cell), only the energy range in the vicinity of the Fermi level is shown. Bands crossing $E_F$ are plotted with color lines. Their number is three in the scalar-relativistic case and four in the full-relativistic case. As can be inferred from the figures, SOC removes band degeneracy in high-symmetry points, see especially T point and L-Γ direction, and significantly influences overall shape of the FS sheets. All parts of FS have a three-dimensional character, except for the FS pockets plotted in panels (c) and (g), which are cylindrical-like in shape with the trigonal axis as their symmetry axis. This feature is a consequence of the presence of atomic layers perpendicular to the trigonal direction, and the largest contribution to that FS sheet comes from metallic layers of Ir(18*h*). It is worth noting that similar FS topology was found before in a sister compound CeIr$_3$ [7].

From the afore-quoted value of DOS at the Fermi level, $N(E_F) = 3.46$ eV$^{-1}$/f.u., one can calculate the Sommerfeld coefficient to be $\gamma_{calc} = 8.155$ mJ/(molK$^2$). This value is distinctly smaller than the experimental one ($\gamma = 17.6$ mJ/(molK$^2$)), and the renormalization factor $\lambda = \gamma/\gamma_{calc} - 1 = 1.16$ is notably larger than the electron-phonon coupling parameter $\lambda_{e-p} = 0.74$ calculated from the McMillan equation (see above). However, if one assumes, that the electronic heat capacity is renormalized by the electron-phonon interaction only, $\lambda = 1.16$ would imply the superconducting critical temperature $T_{c,0} = 10.6$ K, which is much higher than the experimental one. Such discrepancy can be possibly attributed to substantial electron-electron interactions, identified in the low-temperature electrical resistivity data of ThIr$_3$ as the Fermi liquid behavior $\rho \propto AT^2$ with renormalized $A$ coefficient (see above). In line with this hypothesis, similar discrepancy is observed for the magnetic susceptibility values. The computed value of the Pauli paramagnetic susceptibility is $\chi_P = 11 \times 10^{-5}$ emu/mol, thus it is almost four times smaller than the experimental susceptibility $\chi = 40 \times 10^{-5}$ emu/mol. As the experimental susceptibility additionally contains negative diamagnetic contributions, the enhancement of the paramagnetic susceptibility is even larger, validating the picture of important electronic interactions in ThIr$_3$.

To try to quantify this effect, we may follow the route proposed for many d-band superconductors, like Vanadium and its alloys [34–38], Mo$_3$Sb$_7$ [39], La$_3$Co [40] or Y$_3$Rh [41]. Assuming that the electronic correlations take mostly the form of spin fluctuations (electron-paramagnon interactions), characterized by the coupling constant $\lambda_{sf}$, one can compute the superconducting critical temperature from the modified McMillan's formula [39,42]:

$$T_{c,eff} = \frac{\Theta_D}{1.45} \exp\left[\frac{-1.04(1+\lambda_{\text{eff}})}{\lambda_{\text{eff}} - \mu_{eff}(1+0.62\lambda_{\text{eff}})}\right], \quad (9)$$

where $\lambda_{eff} = \frac{\lambda_{e-p}}{1+\lambda_{sf}}$ represents the effective coupling parameter and $\mu_{eff} = \frac{\mu* + \lambda_{sf}}{1+\lambda_{sf}}$ is the effective Coulomb repulsion constant [39]. Assuming the presence of weak electron-paramagnon interactions with postulated $\lambda_{sf} = 0.10$, the electron-phonon coupling parameter and the Coulomb repulsion parameter become renormalized to $\lambda_{eff} = 0.96$ and $\mu_{eff} = 0.21$, yielding $T_{c,eff} = 4.5$ K that is close to the experimental value. Also the renormalized Sommerfeld coefficient $\gamma = \gamma_{calc}(1 + \lambda_{e-p} + \lambda_{sf})$ is now in a good agreement with the experimental value (see **Table 4**).

## 4. Conclusions

The compound ThIr$_3$ crystallizes with a centrosymmetric rhombohedral unit cell of the PuNi$_3$–type, and can be considered as the first actinoid-based representative of the homologous series RE$_{2m+n}$T$_{4m+5n}$, where RE was restricted thus far to a light lanthanide element only, and T = Ir or Rh. The crystal structure is composed from alternating MgCu$_2$– and CaCu$_5$–type blocks. It contains three nonequivalent positions of Ir atoms and two of Th atoms. The characteristic feature is the presence of a layer formed by Ir(6$c$) and Th(3$a$) atoms with fairly short interatomic distance.

As its lanthanide counterparts LaIr$_3$ and CeIr$_3$, ThIr$_3$ becomes superconducting at low temperatures. The bulk nature of the superconducting state in this material is evident from the prominent anomalies at $T_c$ = 4.41 K in its thermodynamic (magnetic susceptibility, heat capacity) and electrical transport characteristics. Our analysis of the experimental and theoretical data revealed that ThIr$_3$ is a d-band, moderately coupled type-II superconductor with isotropic $s$-wave energy gap. The main role in the electronic properties of ThIr$_3$ play the 5$d$-states of iridium with contribution from 6$d$ states of thorium. Calculated occupation of 5f states of Th is comparable to that found in crystalline Th, although these are itinerant and non-magnetic states. At low temperatures, the normal-state electrical conductivity in this material is mostly governed by electron-electron interactions with minor yet not negligible effective mass renormalization. The electronic interactions also additionally renormalize the electronic specific heat coefficient $\gamma$ and the paramagnetic susceptibility, and their competition with the electron-phonon coupling effectively leads to lowering the superconducting critical temperature. Consistent explanation of the magnitude of both $\gamma$ and $T_c$ was reached when a small electron-paramagnon interaction parameter $\lambda_{sf}$ = 0.1 was assumed. It will be worth to investigate other actinoid-based counterparts to examine possible interplay between superconductivity and magnetism in these systems.

**Acknowledgements**

This work was partly supported by the Ministry of Science and Higher Education (Poland) under project DI2016 020546 ("Diamentowy Grant"). Research performed at the AGH-UST was supported by the National Science Center (Poland), project No. 2017/26/E/ST3/00119. SG was partly supported by the EU Project POWR.03.02.00-00-I004/16.

**Figure captions**

**Figure 1.** Powder X-ray diffraction pattern (red points) together with the LeBail refinement profile (black solid line) for ThIr$_3$. The green and orange vertical bars indicate the expected Bragg peak positions for ThIr$_3$ and ThO$_2$ impurity, respectively. Amount of the impurity phase is 1.3(1) wt.%., i.e. 3.9(3) mol%. The blue curve is the difference between the experimental and model results. The inset shows the crystal structure of ThIr$_3$ together with Frank-Kasper polyhedrals with Th atom in the center.

**Figure 2.** Temperature dependencies of the zero-field-cooled (ZFC) and field-cooled (FC) volume magnetic susceptibility measured in a magnetic field of 20 Oe. The raw data were corrected for the demagnetization factor, as described in the main text. The red straight lines illustrate derivation of the critical temperature. The inset shows the isothermal magnetization versus applied magnetic field measured at 2 K with increasing and decreasing field, as indicated by the arrows.

**Figure 3.** (a) Magnetic field dependencies of the magnetization of ThIr$_3$ taken at several different temperatures in the superconducting state upon cooling the sample in zero field. The straight line emphasizes a linear behavior of the isotherm taken at 1.7 K. (b) Analysis of the magnetization isotherms from panel (a), as described in the main text. (c) Temperature variation of the lower critical field derived from panel (b) The red line represents the fit of Eq. 1 to the experimental data.

**Figure 4.** (a) Temperature dependence of the electrical resistivity of ThIr$_3$ measured in zero magnetic field. The inset shows the low-temperature resistivity data taken in several different magnetic fields: 0, 5, 10, 15, 20, 25, 30, 35 and 40 kOe (consecutive curves from right to left). The horizontal dashed line illustrates the criterion used for deriving the critical temperature. (b) Temperature dependence of the upper critical field, determined from the electrical resistivity data. The solid straight line represents the initial slope of the upper critical field curve. (c) Low-temperature electrical resistivity

of ThIr$_3$ measured in a field of 40 kOe. The solid red line emphasizes a Fermi liquid behavior.

**Figure 5.** (a) Temperature dependence of the specific heat over temperature ratio of ThIr$_3$ measured in zero magnetic field in the vicinity of the superconducting phase transition. The thin solid lines illustrate the equal entropy construction used to derive the critical temperature. (b)The specific heat over temperature ratio measured in a magnetic field of 6 T and plotted as a function of squared temperature. The red straight line represents the Debye fit discussed in the main text. (c) Low-temperature variation of the electronic contribution to the zero-field specific heat of ThIr$_3$. The solid red line is the BCS-type fit described in the text.

**Figure 6.** Density of states in ThIr$_3$ calculated without (a) and with (b) spin-orbit coupling effect (solid black lines). The Th and Ir contributions (summed over all atomic positions) are marked by dashed red and dotted green lines, respectively.

**Figure 7.** Comparison of the scalar-relativistic and full-relativistic density of states in ThIr$_3$ near the Fermi level. The results computed without and with spin-orbit coupling effect are shown by dashed red and solid black lines, respectively.

**Figure 8.** Partial density of states in ThIr$_3$ calculated with spin-orbit coupling included. For each Th and Ir atom, the contributions from different orbitals are plotted with different colors. In the upper right corner, two valence electron charge density plots are shown. They are plotted in (010) and (001) planes of the conventional unit cell respectively (on a logarithmic scale, in units of $e/a_B^3$). The Th and Ir atoms are marked with red and green balls respectively. These plots show a metallic character of Ir clusters, an in-plane charge transfer between Th(3a) and its nearest neighbor Ir(6c) and an out-of-plane charge transfer between Th(6c) and Ir(18h) atoms.

**Figure 9.** (a) Brillouin zone used for calculation of the electronic structure of ThIr$_3$. The location of high symmetry points is indicated. (b-d) Fermi surface pockets calculated in scalar-relativistic approach. (e-h) Fermi surface calculated with spin-orbit coupling effect included. (i) Band structure in the vicinity of the Fermi level calculated without spin-orbit coupling. (j) Band structure computed in full-relativistic approach.

**Table 1**

Crystallographic data for ThIr$_3$. The reliability factors provided are conventional Rietveld *R*-factors (corrected for background contribution) calculated only for points with Bragg contribution. $B_{iso}$ stands for equivalent isotropic displacement parameters. Numbers in parentheses are statistical uncertainties of the least significant digits and are not corrected for the presence of possible experimental errors.

| Space group | R-3m (#166) | | Reliability factors | |
|---|---|---|---|---|
| *a* (Å) | 5.3397(1) | | $R_p$ (%) | 15.4 |
| *c* (Å) | 26.4213(8) | | $R_{wp}$ (%) | 15.5 |
| *V* (Å$^3$) | 653.09 | | $R_{exp}$ (%) | 6.09 |
| Density(g/cm$^3$) | 18.5 | | $\chi^2$ | 6.51 |
| Atom (position): | x | y | Z | $B_{iso}$ (Å$^2$) |
| Th1 (3*a*) | 0 | 0 | 0 | 0.85(7) |
| Th2 (6*c*) | 0 | 0 | 0.1396(1) | 1.56(6) |
| Ir1 (3*b*) | 0 | 0 | ½ | 0.87(8) |
| Ir2 (6*c*) | 0 | 0 | ⅓ | 0.50(6) |
| Ir3 (18*h*) | 0.4995(2) | 0.5005(2) | 0.0829 | 0.75(3) |

**Table 2**

Superconducting and normal state parameters of ThIr$_3$, CeIr$_3$ [7] and LaIr$_3$ [5].

| Parameter | Unit | ThIr$_3$ | CeIr$_3$ | LaIr$_3$ |
|---|---|---|---|---|
| $T_c$ | K | 4.41 | 2.46 | 3.32 |
| $H_{c2}(0)$ | kOe | 47 | 35 | 38.4 |
| $H_{c1}(0)$ | Oe | 60 | 173 | 1102 |
| $H_c$ | Oe | 884 | 1470 | 1750 |
| $\lambda_{e\text{-}p}$ | --- | 0.74 | 0.65 | 0.57 |
| $\xi_{GL}(0)$ | Å | 83 | 92 | 92.59 |
| $\lambda_{GL}(0)$ | Å | 3150 | 1640 | 960 |
| $\kappa$ | --- | 38 | 17 | 10.37 |
| $\gamma$ | mJ/(mol K$^2$) | 17.6 | 25.1 | 11.5 |
| $\beta$ | mJ/(mol K$^4$) | 1.59 | 2.72 | --- |
| $\Theta_D$ | K | 169 | 142 | 366 |
| RRR | --- | 1.6 | 1.7 | --- |
| $\Delta C/\gamma T_c$ | --- | 1.6 | 1.24 | 1.22 |

**Table 3**
Orbital filling in ThIr$_3$ calculated as an integral of DOS over energy, using DOS from scalar-relativistic and full-relativistic calculations. Partial DOS at the Fermi level contributed by each atom is also given (in eV$^{-1}$).

|  | w/o SOC | | | | | with SOC | | | | |
|---|---|---|---|---|---|---|---|---|---|---|
|  | Ir(3$b$) | Ir(6$c$) | Ir(18$h$) | Th(3$a$) | Th(6$c$) | Ir(3$b$) | Ir(6$c$) | Ir(18$h$) | Th(3$a$) | Th(6$c$) |
| No. of valence electrons | 9 | 9 | 9 | 4 | 4 | 9 | 9 | 9 | 4 | 4 |
| charge $Q$(e) | 8.57 | 8.68 | 8.64 | 4.58 | 4.61 | 8.53 | 8.64 | 8.60 | 4.68 | 4.69 |
| $Q$ of s-states | 0.73 | 0.80 | 0.80 | 0.63 | 0.59 | 0.71 | 0.79 | 0.79 | 0.65 | 0.59 |
| $Q$ of p-states | 0.04 | 0.03 | 0.04 | 0.22 | 0.26 | 0.04 | 0.03 | 0.04 | 0.25 | 0.28 |
| $Q$ of d-states | 7.80 | 7.84 | 7.80 | 2.96 | 2.95 | 7.78 | 7.82 | 7.77 | 3.00 | 2.98 |
| $Q$ of f-states |  |  |  | 0.76 | 0.81 |  |  |  | 0.79 | 0.84 |
| DOS at $E_F$ | 0.47 | 0.71 | 0.61 | 0.67 | 0.44 | 0.65 | 1.29 | 0.77 | 0.93 | 0.62 |
| Total DOS at $E_F$ | 2.45 (eV$^{-1}$/f.u.) | | | | | 3.46 (eV$^{-1}$/f.u.) | | | | |

**Table 4**

Analysis of the electron-phonon interaction and the superconducting critical temperature of ThIr$_3$ assuming the presence of effective depairing electron-paramagnon interactions. $T_{c,0}$ is the critical temperature calculated assuming renormalization due to electron-phonon interaction, and $T_{c,\text{eff}}$ is the effective critical temperature computed with addition of spin fluctuations effect.

For the meaning of all the other symbols see the main text.

| Parameter | Unit | Value |
|---|---|---|
| $\Theta_D$ | K | 169 |
| $\gamma$ | mJ/(mol K$^2$) | 17.6 |
| $\gamma_{\text{calc}}$ | mJ/(mol K$^2$) | 8.155 |
| $\lambda = \gamma/\gamma_{\text{calc}} - 1$ | – | 1.158 |
| $T_{c,0}$ ($\mu^*=0.13$) | K | 10.63 |
| $\lambda_{\text{e-p}}$ | – | 0.74 |
| $\lambda_{\text{sf}}$ | – | 0.1 |
| $\mu_{eff}$ | – | 0.209 |
| $\lambda_{\text{eff}}$ | – | 0.962 |
| $T_{c,\text{eff}}$ | K | 4.55 |
| $T_c$ | K | 4.41 |

**Figure 1**

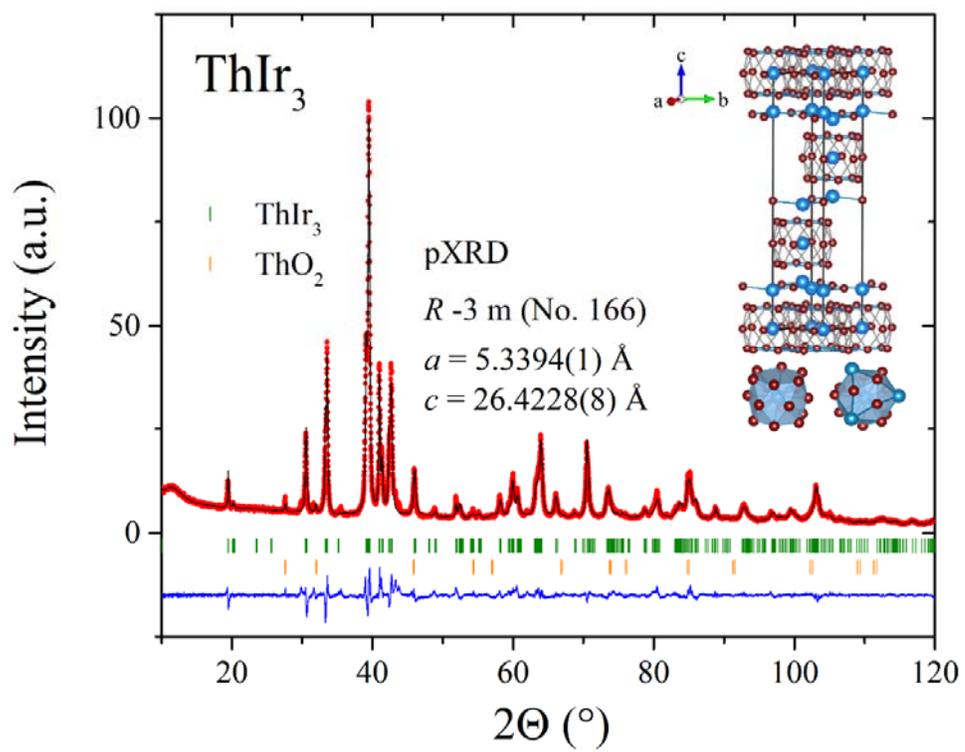

**Figure 2**

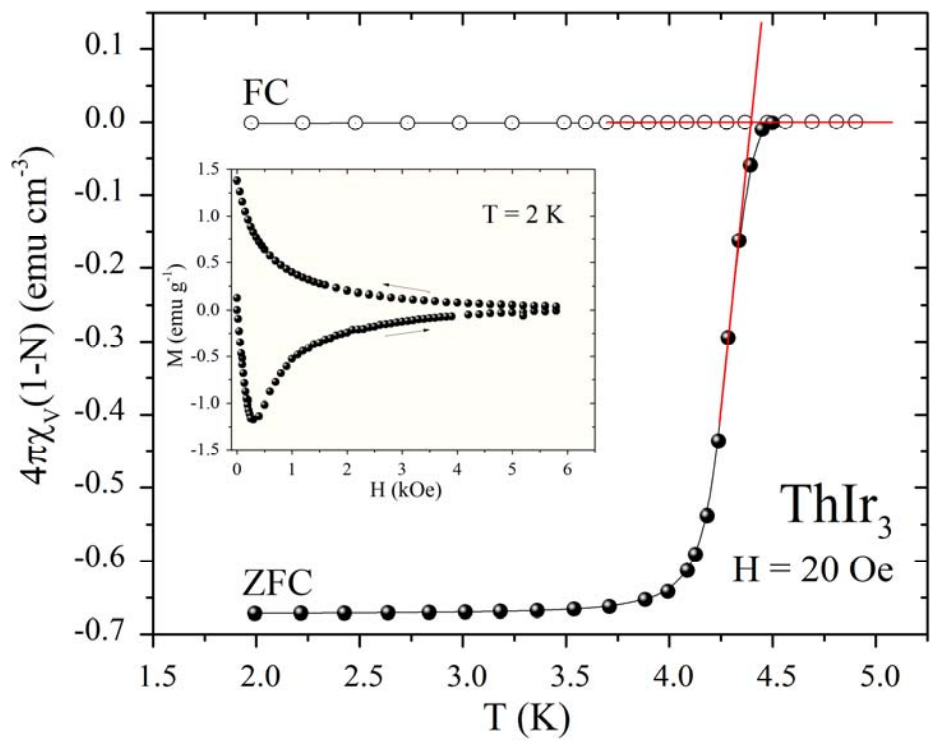

**Figure 3**

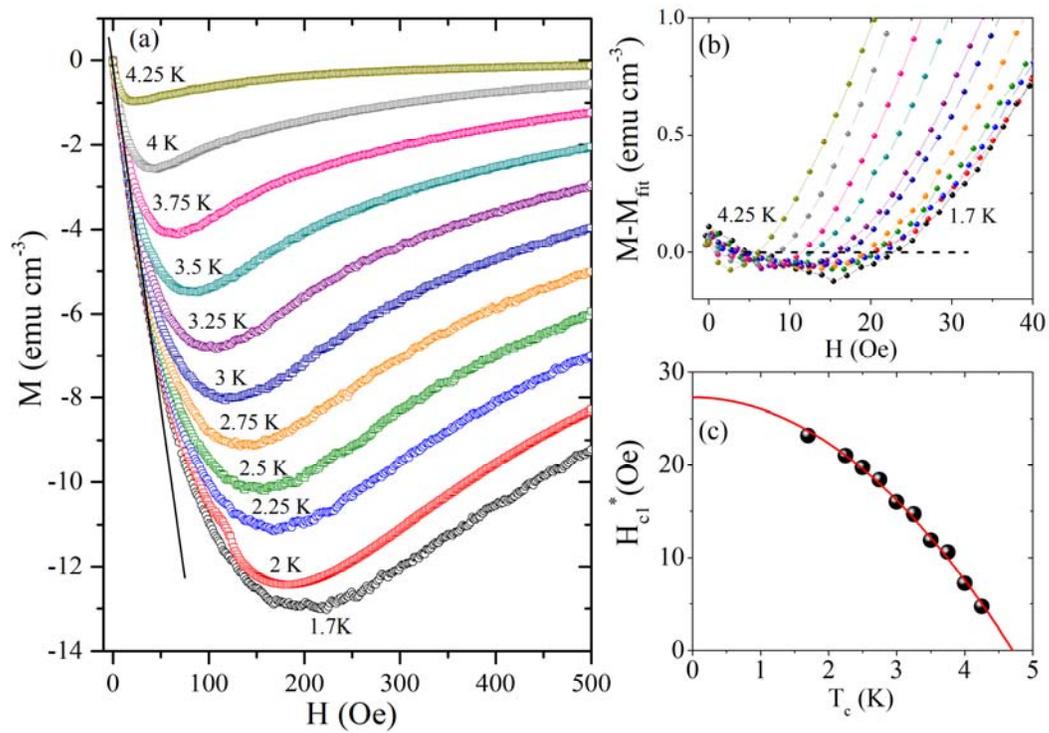

**Figure 4**

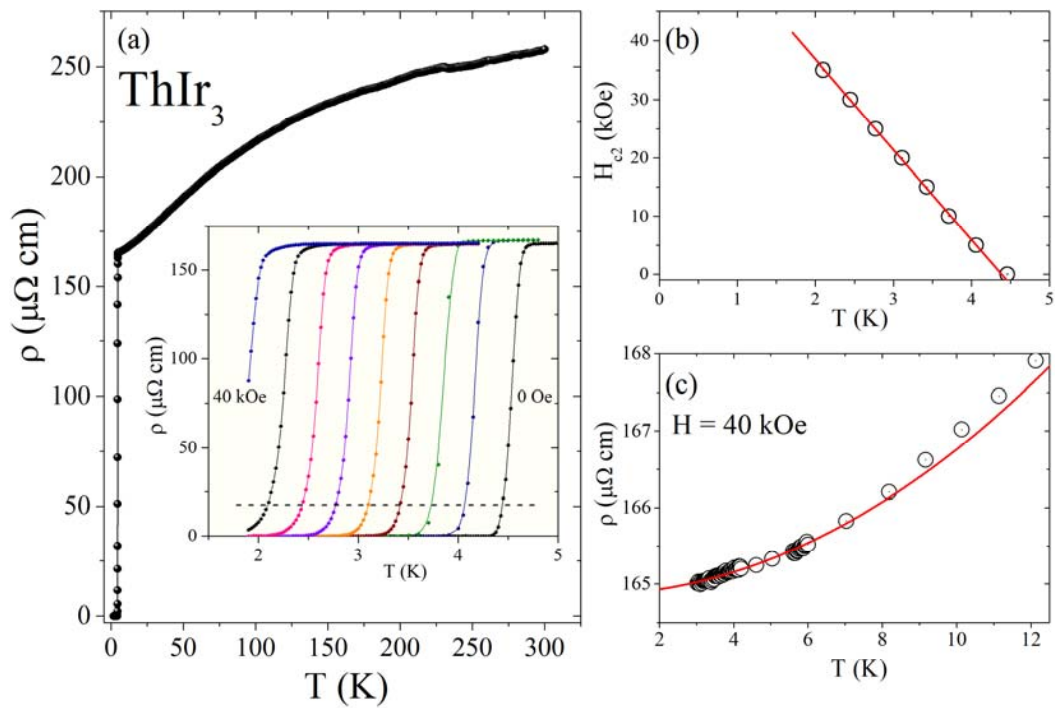

**Figure 5**

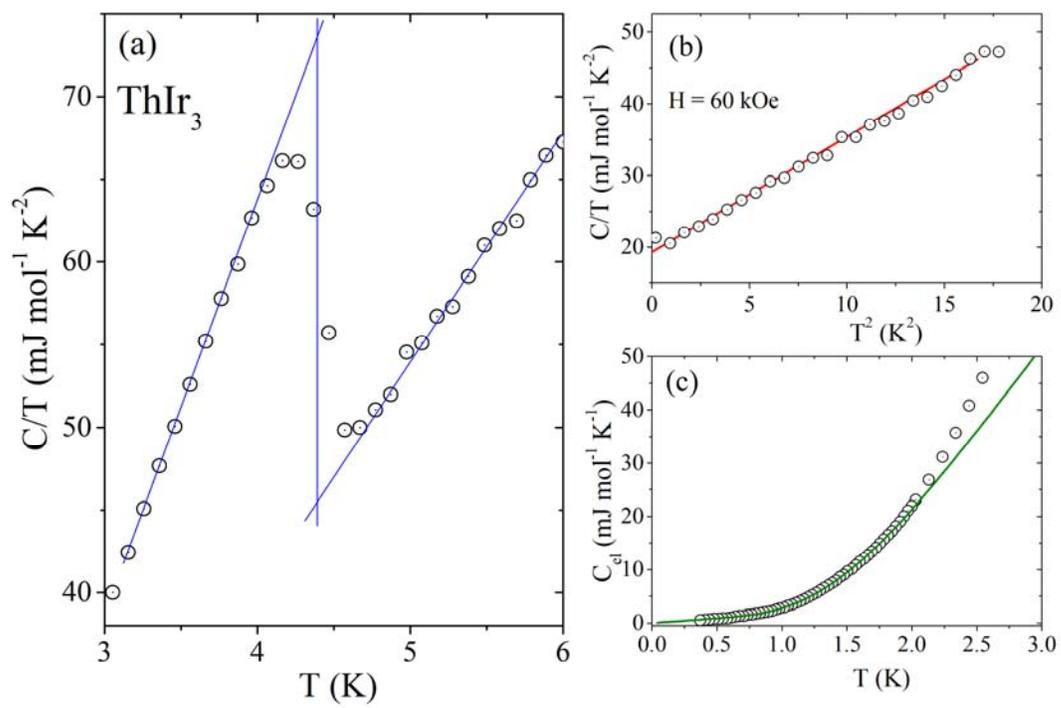

**Figure 6**

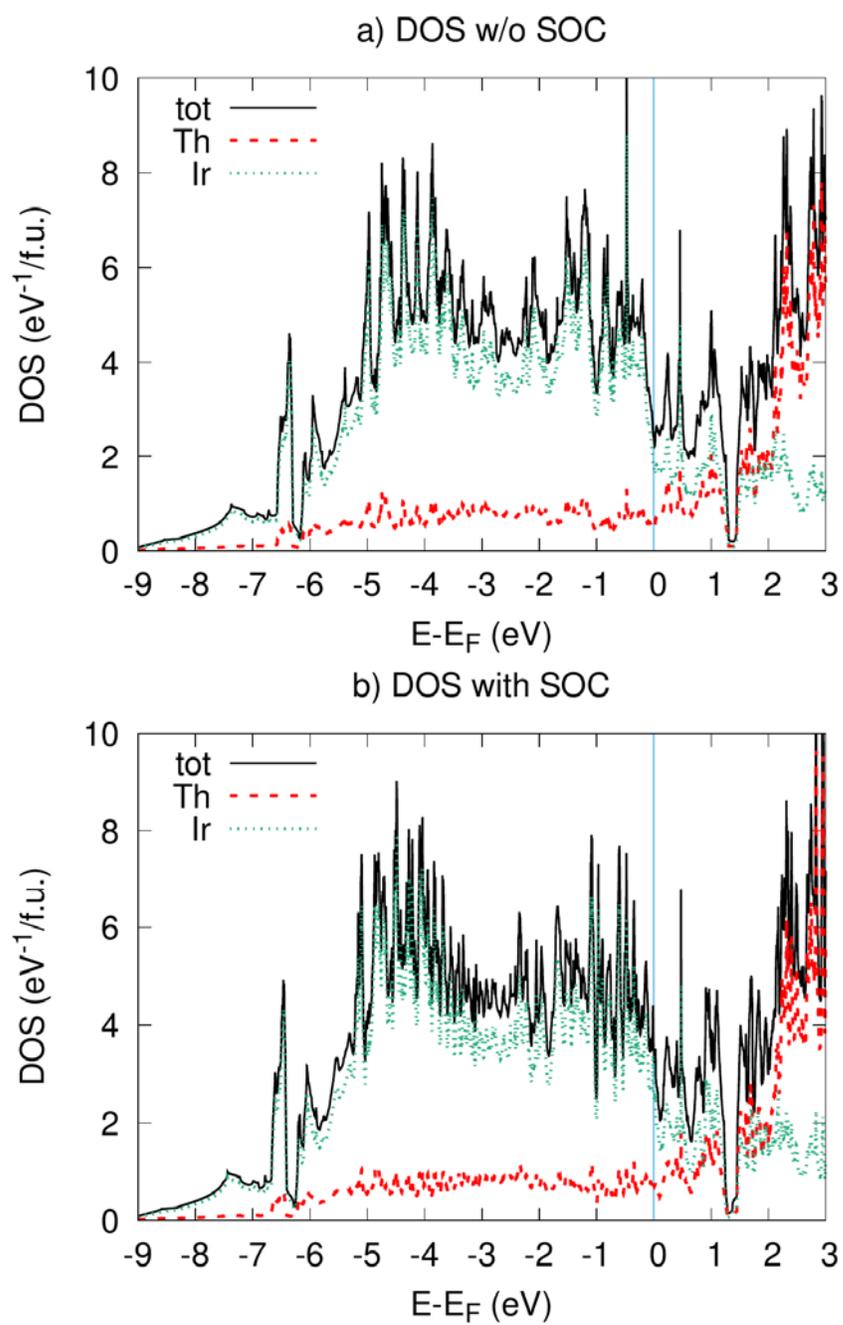

**Figure 7**

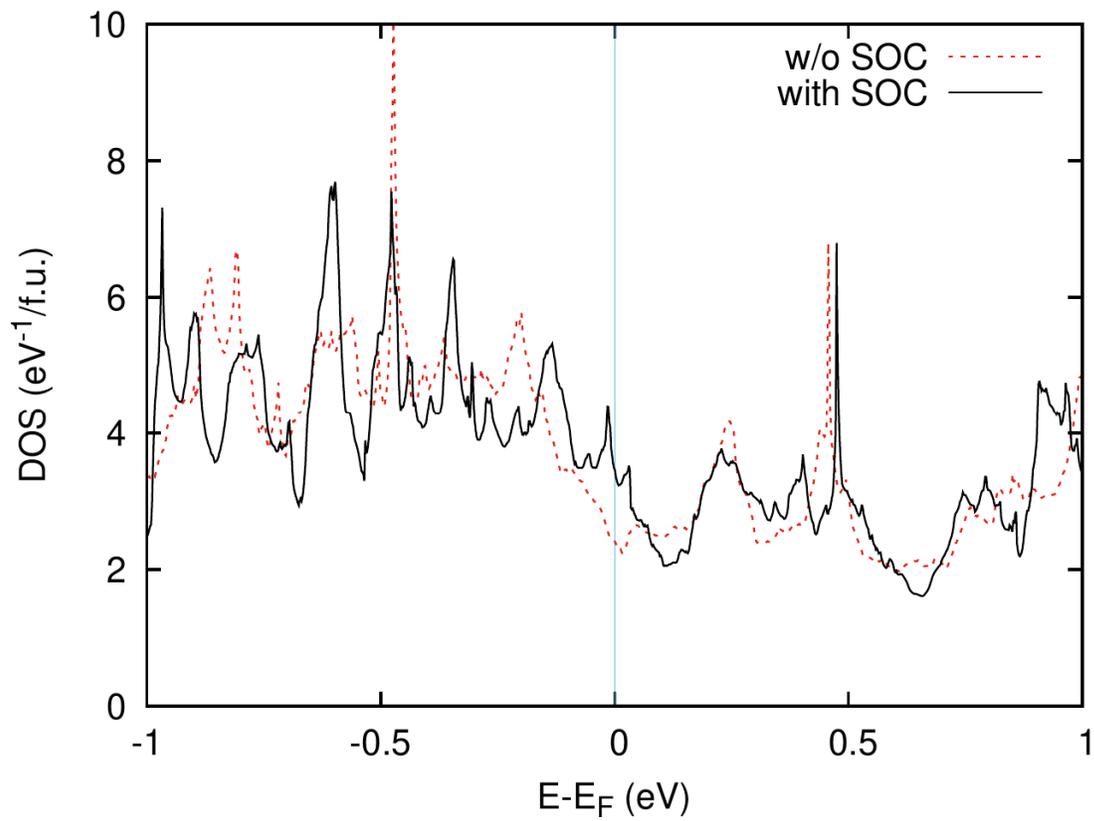

**Figure 8**

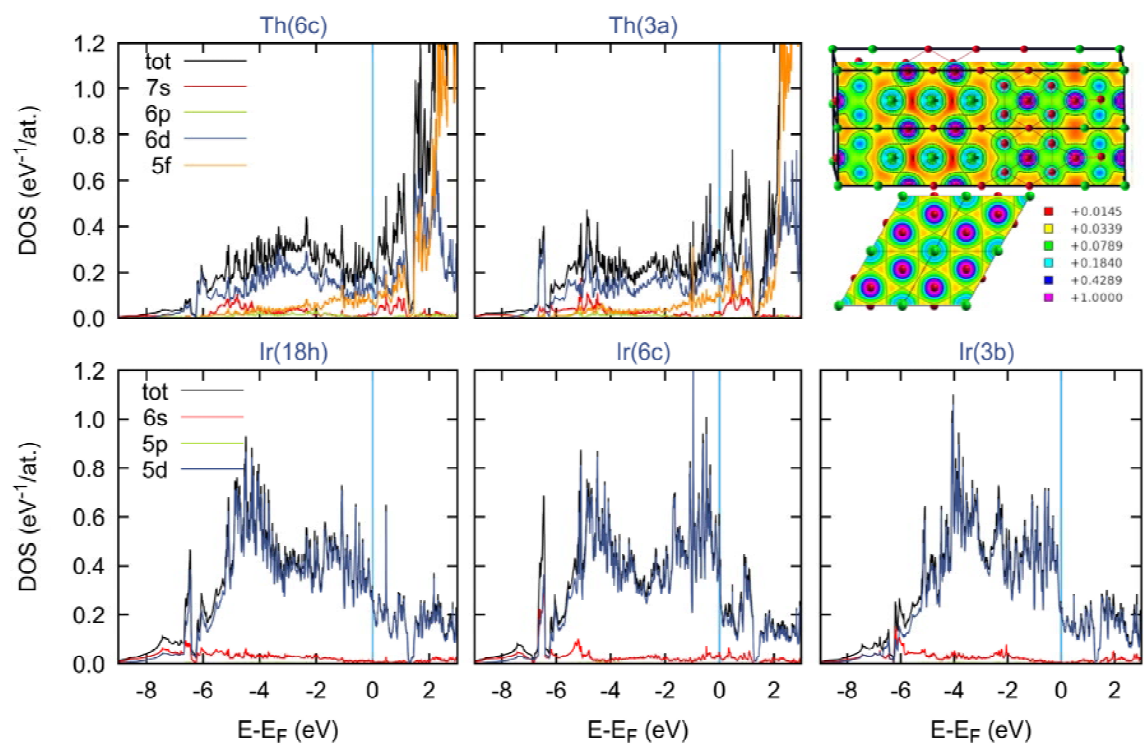

**Figure 9**

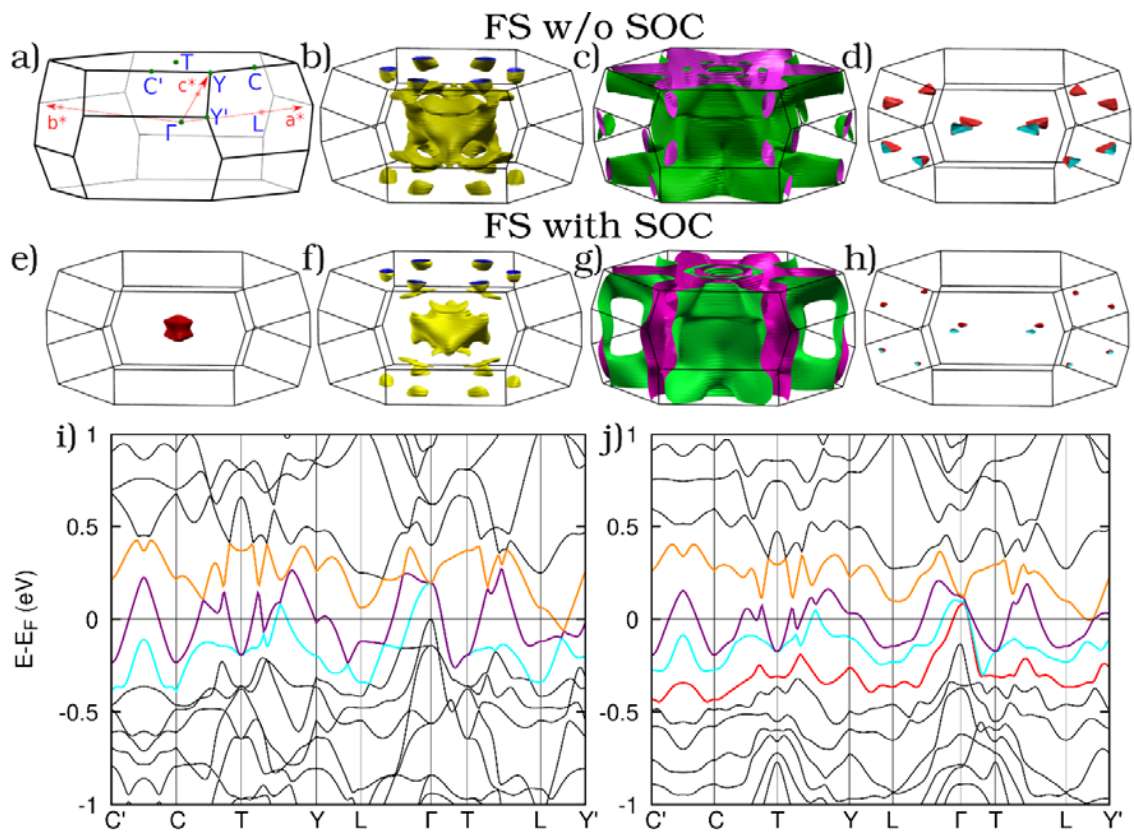